\documentclass[11pt]{article}

\usepackage[final]{acl}

\usepackage{times}
\usepackage{latexsym}

\usepackage[T1]{fontenc}

\usepackage[utf8]{inputenc}

\usepackage{microtype}

\usepackage{inconsolata}

\usepackage{graphicx}
\usepackage{booktabs}
\usepackage{multirow}
\usepackage{makecell}
\usepackage{subcaption}
\usepackage{xcolor}
\usepackage[table]{xcolor}
\usepackage{tabularx}
\usepackage{amsmath}
\title{QReason: Query-Focused Decoupled Chain-of-Thought for Efficient Passage Reranking}

\author{
\textbf{Yang Zhang\textsuperscript{1,2,3}},
\textbf{Wenhan Liu\textsuperscript{4}},
\textbf{Qiannan Zhu\textsuperscript{1,2,3}}\thanks{Corresponding author.},
\textbf{Mingming Li\textsuperscript{5}},
\textbf{Yuanfei Huang\textsuperscript{1,2,3}}
\\
\textsuperscript{1}School of Artificial Intelligence, Beijing Normal University \\
\textsuperscript{2}Beijing Key Laboratory of Artificial Intelligence for Education \\
\textsuperscript{3}Engineering Research Center of Intelligent Technology and Educational Application, \\
Ministry of Education \\
\textsuperscript{4}Gaoling School of Artificial Intelligence, Renmin University of China \\
\textsuperscript{5}Institute of Information Engineering, Chinese Academy of Sciences \\
\texttt{yangzsat@mail.bnu.edu.cn}
}

\begin{document}
\maketitle
\begin{abstract}
Passage reranking plays a crucial role in information retrieval by refining the ordering of candidate passages to better reflect relevance. Existing listwise LLM rerankers with Chain-of-Thought (CoT) reasoning can handle complex queries effectively, but they suffer from substantial redundancy and high latency due to sliding-window strategies, which repeatedly generate highly similar CoTs. 
To address this, we propose QReason, a decoupled framework that separates query-focused reasoning from window-specific passage relevance assessment. Specifically, QReason introduces a dedicated rewriter that generates a ranking-oriented reasoning query once, capturing the query's core intent while avoiding redundant reasoning, and then reuses it across all windows with a non-reasoning reranker. The rewriter is trained via a two-stage process that first uses supervised fine-tuning with relevant-passage guidance through semantic evidence to produce deeply grounded, query-focused CoTs. It then applies reinforcement learning to align CoT generation with both the inference-time setting and the reranking objective, optimizing listwise metrics and passage-level discrimination to produce reusable reasoning chains for reranking. Experiments on the BRIGHT benchmark demonstrate that QReason significantly reduces redundant reasoning, achieves ranking performance comparable to or better than strong reasoning-based rerankers, and outperforms existing query rewriting models.
\end{abstract}

\section{Introduction}
Passage reranking is a critical stage in Information Retrieval, aiming to produce a finer-grained ordering of retrieved passages. Listwise reranking methods based on Large Language Models (LLMs) \citep{rankgpt,zeroshotlistwise,rankzephyr,rankvicuna,coranking,sliding-win-r-not-the-end} have attracted substantial attention because they are able to model relative relationships among passages for accurate ranking. More recently, researchers have further introduced the reasoning capability of LLMs into reranking \citep{rank1,rankk,rankr1,reasonrank_arxiv}, explicitly generating CoT \cite{CoT} during the ranking process to handle relevance judgment for complex queries. However, due to the limitation of context window size, these methods generally rely on sliding-window strategies \cite{rankgpt}, which require repeated CoT generation for the same query across multiple windows, thereby causing substantial computational redundancy and increased response latency.
\begin{figure}[t]
  \centering
  \begin{subfigure}[t]{\columnwidth}
    \centering
    \includegraphics[width=\linewidth]{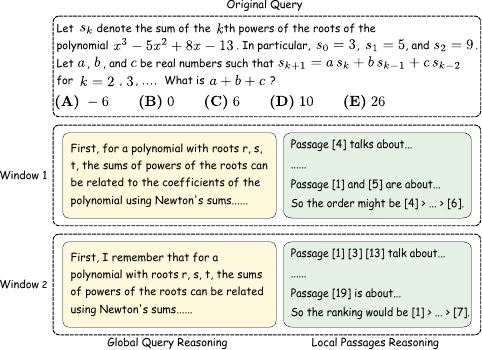}
    \caption{An example of query CoT redundancy across windows.}
    \label{fig:wins_overlap_a}
  \end{subfigure}

  \vspace{0.5em}

  \begin{subfigure}[t]{\columnwidth}
    \centering
    \includegraphics[width=\linewidth]{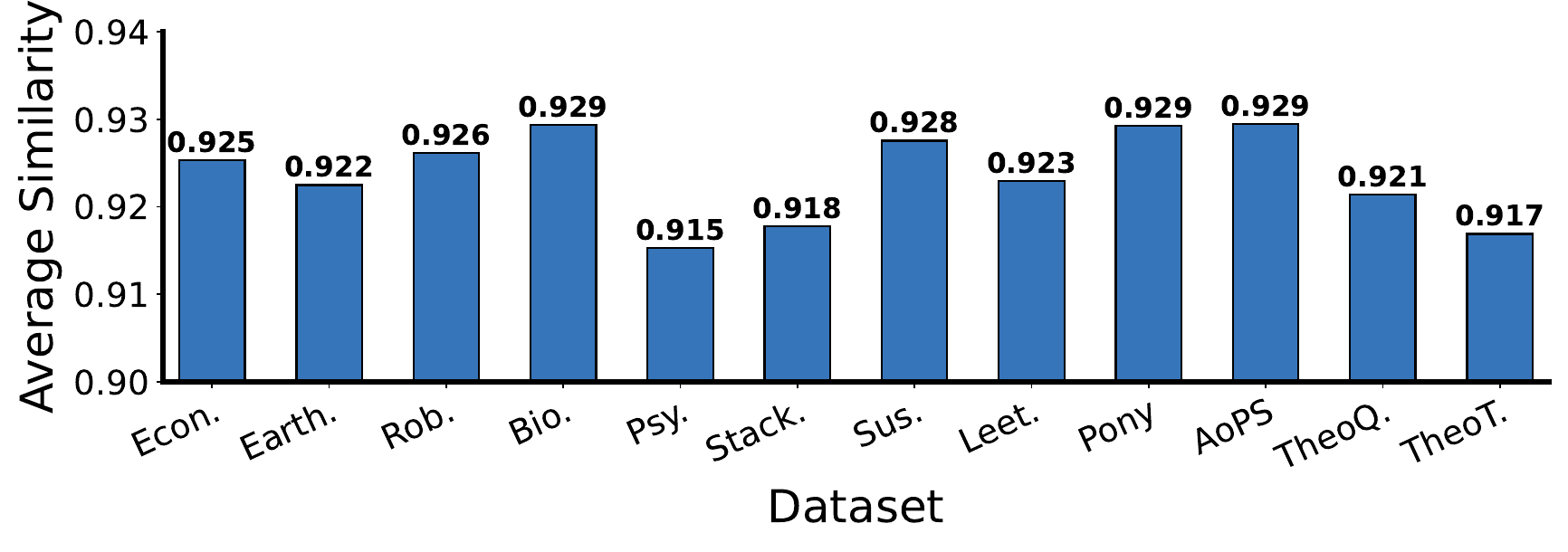}
    \caption{Similarity of query CoTs among windows across datasets.}
    \label{fig:avg_sim_b}
  \end{subfigure}

  \caption{Illustration of CoT redundancy in listwise rerankers.}
  \label{fig:cot_redundancy}
\end{figure}

We conduct an analysis of existing approaches to identify CoT redundancy, and observe that the generated reasoning traces generally consist of two parts: global query reasoning and local passages reasoning, as illustrated in Figure \ref{fig:cot_redundancy}(a). The query reasoning focuses on interpreting the core intent of the query and inferring the characteristics of potentially relevant passages in the global view. The passages reasoning analyzes the passages within the current window and determines whether they are relevant to the query. The former is independent of the specific passages in each window and already captures general reasoning about the characteristics of relevant passages, yet existing methods still regenerate the full reasoning trace for each window, resulting in substantial redundancy. To better quantify this phenomenon, we further analyze the reasoning traces produced by the representative reasoning-based reranking method ReasonRank \cite{reasonrank_arxiv} on the BRIGHT benchmark \cite{bright}, and find that, for the same query, the average semantic similarity between reasoning traces generated in different windows exceeds 90\%, as measured using BGE-m3 embeddings \cite{bgem3}, as shown in Fig.~\ref{fig:cot_redundancy}(b). This high similarity indicates that most of the reasoning trace is repetitive and could, in principle, be reused across windows rather than recomputed.



Motivated by these insights, we propose QReason, a decoupled framework separating global query reasoning from window-specific reasoning by rewriting the original query into a reasoning query, i.e., a CoT-style reformulation of the query. The generated reasoning query is subsequently reused across windows by a non-reasoning reranker to eliminate redundant reasoning while preserving strong ranking performance. Therefore, our core objective is to train a rewriter that transforms the original query into a ranking-oriented reasoning query. To this end, we introduce a two-stage optimization framework. In the first stage, the rewriter is fine-tuned on the pairs of queries and their gold relevant-passages, enabling it to better capture query intent and generate grounded reasoning queries. In the second stage, the rewriter is further refined through reinforcement learning with two complementary rewards: a listwise reward that directly aligns rewriting with reranking quality, and a pointwise reward that constrains the generated reasoning query to remain supported by gold relevant-passages. Together, the two-stage optimization enables the rewriter to produce a query-focused and reusable reasoning query that efficiently guides reranking across candidate passages. Experiments on real-world datasets show that QReason consistently improves non-reasoning rerankers, and its strongest instantiation with Qwen3.5-35B-A3B \cite{qwen3_5} achieves superior ranking accuracy and computational efficiency compared with state-of-the-art reasoning-based rerankers.

In summary, our paper has three contributions: 
\begin{itemize}
  \item \textbf{Analysis of CoT Redundancy}. We reveal that reasoning-aware reranking generates highly similar query-level Chain-of-Thought across sliding windows, introducing substantial redundant reasoning that can be reused instead of repeatedly recomputed.
  \item \textbf{Novel QReason Framework}. We propose a decoupled architecture integrating a query-focused rewriter and a lightweight non-reasoning reranker, which reduces computational overhead while preserving strong ranking performance.
  \item \textbf{Two-Stage Rewriter Optimization}. We optimize the rewriter via supervised fine-tuning followed by reinforcement learning, yielding accurate, ranking-aligned reasoning queries for efficient and effective reranking.
\end{itemize}

\section{Related Work}
\textbf{Reasoning-Intensive LLM Reranking}\quad In recent years, large language models have been increasingly adopted for reranking in information retrieval due to their strong semantic modeling ability \citep{rankt5,rankzephyr,rankvicuna,pairwise-ranking,llm4ir,rankllm,demorank}. Nevertheless, prior studies \cite{bright} have shown that existing LLM rerankers still perform poorly on reasoning-intensive IR tasks. Unlike conventional IR tasks that mainly rely on keyword and semantic matching, reasoning-intensive IR tasks require deeper query understanding and multi-step reasoning. To address this challenge, recent work has incorporated explicit reasoning into the reranking process. For instance, Rank-K \cite{rankk} and Rank1 \cite{rank1} distill reasoning traces to supervise reranker fine-tuning, while ReasonRank \cite{reasonrank_arxiv}, Rank-r1 \cite{rankr1}, and REARANK \cite{rearank} optimize rerankers using GRPO \cite{grpo}. Although these approaches improve effectiveness on reasoning-intensive IR tasks, they typically require the model to explicitly generate reasoning traces during inference, resulting in considerable computational cost and substantial redundancy. 

\noindent\textbf{Reasoning-Enhanced Query Rewriting} \quad Query rewriting has been widely adopted to improve first-stage retrieval by reformulating under-specified or ambiguous queries \citep{qesurvey,query2doc,queryexpansion,PRF}. Recently, researchers have begun to leverage the reasoning capabilities of LLMs to address complex IR tasks. For example, some methods \cite{bright} directly prompt large-scale LLMs, such as GPT-4 \cite{gpt4} or LLaMA3-70B-Instruct \cite{llama3}, to generate reasoning-based rewrites, while others train smaller rewriters, such as DeepRetrieval \cite{deepretrieval} and TongSearch \cite{tongsearch}, using reinforcement learning to endow them with reasoning capabilities. Although these approaches have demonstrated strong effectiveness for retrieval, they are primarily designed for the retrieval stage and do not transfer well to reranking.

\section{Task Formulation}
Given a query $q$ and candidate passages $\mathcal{P}$, listwise reranking under context-length constraints divides $\mathcal{P}$ into overlapping windows $\mathcal{W}=\{W_1,\dots,W_K\}$. Existing reasoning-based rerankers process each window independently:
\begin{equation}
c_j = g(q,W_j), \quad \pi_j = f(q,W_j,c_j),
\end{equation}
where a new reasoning trace $c_j$ is independently generated for each window $W_j$, and $\pi_j$ denotes the permutation produced by reranker $f$, ordering the passages in $W_j$ by their estimated relevance to the query. This leads to substantial redundancy.

QReason instead decouples reusable query reasoning from window-level ranking:
\begin{equation}
q'=\mathcal{RW}(q), \quad \pi_j=f(q,W_j,q').
\end{equation}
Here, the rewriter generates a reranking-oriented reasoning query $q'$ once, which is then reused across all windows by a non-reasoning reranker.
 




\section{Method}

\begin{figure*}[t]
    \centering
    \includegraphics[width=\textwidth]{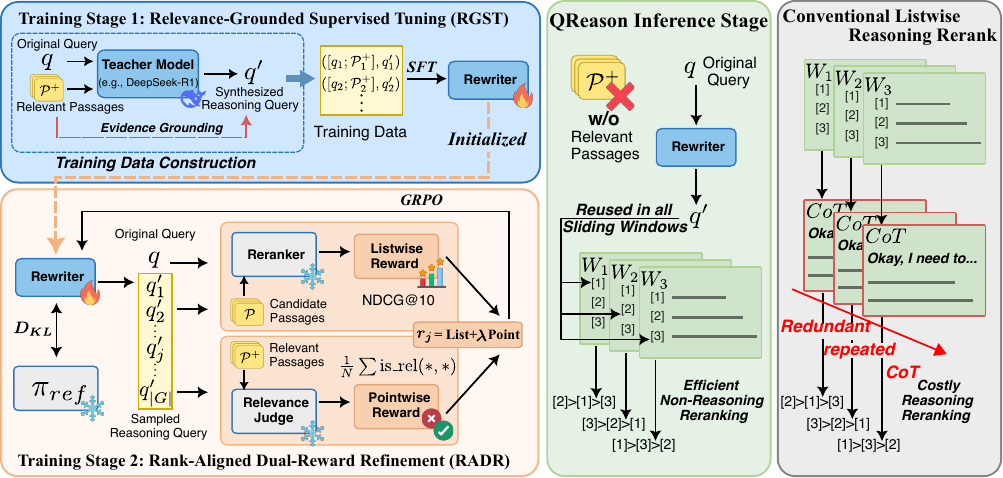}
    \caption{Overall framework of QReason, including two-stage training and inference-time comparison with conventional listwise reasoning reranker.}
    \label{fig:method}
\end{figure*}


As discussed above, an effective way to reduce redundancy is to externalize global query reasoning as a query rewrite that is independent of the local candidate passages, so that it can be reused across sliding windows without repeated generation. Though many retrieval-stage query rewriting methods notably boost retrieval performance, they cannot be directly adopted for reranking. Motivated by \citet{qe4Cross-Encoder-Rankers}, we argue that rewriting for reranking differs from rewriting for retrieval in both objectives and constraints. Since rerankers jointly process the query and documents under a precision-oriented objective, they are more sensitive to fine-grained query-document interactions and to additional content introduced by rewriting. Additional content that may be harmless in retrieval can therefore deviate from the core ranking intent, causing false relevance signals and impairing ranking performance. Our experiments show that, although rewrites from GPT-4 and TongSearch are effective for retrieval, their effects become unstable when applied to reranking.

Building on this insight, we propose QReason, a decoupled framework that extracts global query reasoning by a \textit{rewriter} as a CoT-style reasoning query and reuses it across windows in the non-reasoning reranker to eliminate redundant computation. The \textit{rewriter} is first trained via supervised fine-tuning on (query, relevant-passages) pairs to capture intent accurately, and then refined with reinforcement learning using both listwise and pointwise rewards to align with inference-time setting and reranking objectives.

\subsection{Relevance-Grounded Supervised Tuning}
The rewriter in QReason is expected to generate a reasoning query that captures the underlying intent and integrates relevant passage evidence. To realize this goal, we argue that gold relevant-passages serve as the most reliable supervision signal, since they clearly define the critical query facets required for accurate relevance judgment. Accordingly, by introducing gold relevant-passages into the Relevance-Grounded Supervised Tuning (RGST) stage, we reformulate the rewriting task from an unconstrained query expansion into a grounded reasoning task guided by relevance information.

Formally, given a query $q$ and its gold relevant-passages $\mathcal{P^{+}}$, we leverage a powerful teacher model (e.g., DeepSeek-R1) to synthesize the reasoning query $q'$, which is specifically prompted to distill the reasoning chain connecting $q$ to $\mathcal{P^{+}}$, highlighting the core intent required for precise passage ranking. Using the synthesized reasoning queries, 
we employ supervised fine-tuning (SFT) to train the rewriter, where each training sample is an input-output pair as $([q; \mathcal{P^{+}}], q')$.
The optimization goal is the standard next-token prediction objective:
\begin{equation}
\mathcal{L}_{\mathrm{RGST}}
=
-\sum_{t=1}^{|q'|}
\log p_{\theta}\!\left(q'_t \mid q, \mathcal{P^{+}}, q'_{<t}\right),
\label{eq:L_RGST}
\end{equation}
where $p_{\theta}$ represents the rewriter parameterized by $\theta$.


\subsection{Rank-Aligned Dual-Reward Refinement}
Although RGST guides the rewriter to generate evidence-grounded reasoning queries that capture key relevance patterns, there remains a training--inference discrepancy, as $\mathcal{P}^{+}$ is used during training but unavailable at inference time. Moreover, RGST does not directly optimize downstream reranking performance. QReason addresses these limitations through Rank-Aligned Dual-Reward Refinement (RADR), which introduces a reinforcement learning framework with task-specific rewards, including listwise rewards for overall ranking quality and pointwise rewards for passage-level discrimination. This stage directly aligns query generation with both the inference-time setting and the reranking objective, encouraging the rewriter to produce queries that are grounded and effective for precise ranking.

For each query $q$, we sample a group of $|G|$ reasoning queries from the current rewriter policy:
\[
\mathcal{G}(q)=\{q'_1,q'_2,\dots,q'_{|G|}\}, \qquad q'_j \sim \pi_{\theta}(\cdot \mid q).
\]
Each sampled reasoning query $q'_j$ is then evaluated by rewards tailored to reranking.

\noindent\textbf{Listwise reward.} The listwise reward aligns query generation with the downstream reranking objective. Specifically, for each sampled reasoning query $q'_j$ in $\mathcal{G}(q)$, we feed the original query $q$, the rewritten reasoning query $q'_j$, and the candidate passages into a fixed \textit{reranker} to obtain a reranked permutation $\pi_j$. The listwise reward is given by the NDCG@10 score of the resulting ranking, i.e.,
\[
\mathrm{List}(q'_j)=\mathrm{NDCG@10}(\pi_j,\mathcal{Y}_q),
\]
where $\mathcal{Y}_q$ denotes the relevance labels of the candidate passages for query $q$.  

\noindent\textbf{Pointwise reward.}
Optimizing only the listwise signal may cause the rewriter to overfit to the preference of a single fixed reranker, especially when that reranker has limited ranking capacity. To prevent the generated reasoning query from drifting away from the true semantic scope of relevant evidence, we introduce a pointwise reward based on gold relevant-passages, which constrains the rewrite to remain evidence-supported and avoids degenerate generations that improve reranker scores while being weakly related to actual relevance evidence.


Let $\mathcal{P}^{+}(q)=\{p_{q,i}^{+}\}_{i=1}^{N_q}$ denote the set of gold relevant passages for query $q$, where $N_q$ denotes the number of judged relevant passages for $q$. For each sampled reasoning query $q'_j$, we use a \textit{relevance judge} to independently determine whether $q'_j$ is semantically consistent with each $p_i^{+}$. The pointwise reward is defined as
\[
\mathrm{Point}(q'_j)
=
\frac{1}{N_q}\sum_{i=1}^{N_q}\mathrm{is\_rel}(q'_j,p_{q,i}^{+}),
\]
where $\mathrm{is\_rel}(q'_j,p_{q,i}^{+})\in\{0,1\}$ is a binary judgment function that returns $1$ if the judge considers $q'_j$ relevant to $p_{q,i}^{+}$, and $0$ otherwise. 

The final reward for each sampled reasoning query $q'_j$ is defined as
\[
r_j
=
\mathrm{List}(q'_j)+\lambda\,\mathrm{Point}(q'_j),
\]
where $\lambda$ controls the trade-off between direct ranking optimization and relevance-grounded semantic constraint.

Following GRPO \cite{grpo}, rewards are normalized within each sampled group to obtain the relative advantage $\hat{A}_j$, which is then used to optimize the rewriter with the clipped objective and KL regularization:
\begin{equation}
\begin{aligned}
\mathcal{L}_{\mathrm{RADR}}(\theta)
=&-\frac{1}{|G|}
\sum_{j=1}^{|G|}
\frac{1}{|q'_j|}
\sum_{t=1}^{|q'_j|}
\Bigg[
\min\Big(
\rho_{j,t}(\theta)\hat{A}_j, \\
&\hspace{-1.8em}
\mathrm{clip}\!\big(
\rho_{j,t}(\theta),\,1-\varepsilon,\,1+\varepsilon
\big)\hat{A}_j
\Big) \\
&\hspace{-1.8em}
-\beta D_{\mathrm{KL}}\!\Big(
\pi_{\theta}(\cdot\mid s_{j,t})
\,\|\,
\pi_{\mathrm{ref}}(\cdot\mid s_{j,t})
\Big)
\Bigg], \\
\rho_{j,t}(\theta)
=&\frac{\pi_{\theta}(q'_{j,t}\mid s_{j,t})}
{\pi_{\theta_{\mathrm{old}}}(q'_{j,t}\mid s_{j,t})}.
\end{aligned}
\label{eq:L_RADR}
\end{equation}

\begin{table*}[t]
\centering
\caption{Performance comparison on BRIGHT. The best score is shown in \textbf{bold} and the second best is \underline{underlined}. Rows denoted as ``with X'' use queries from X for reranking, where X can be the original queries or rewritten queries generated by GPT-4, DeepRetrieval, TongSearch, or QReason. All reranker baseline results are reproduced by us under the same experimental environment.}

\scriptsize
\setlength{\tabcolsep}{4pt}
\renewcommand{\arraystretch}{1.08}
\begin{tabular}{lccccccccccccc}
\toprule
 & Econ. & Earth. & Rob. & Bio. & Psy. & Stack. & Sus. & Leet. & Pony & AoPS & TheoQ. & TheoT. & Avg \\
\midrule

ReasonIR
& 32.43 & 44.19 & 20.82 & 43.49 & 39.70 & 30.96 & 27.34 & \textbf{32.33} & 19.23 & 8.14 & 32.84 & 36.41 & 30.66 \\
\midrule

\multicolumn{14}{c}{{\small\textbf{\textit{Task-Specific Rerankers}}}} \\
\midrule
RankT5 (3B)
& 11.28 & 22.17 & 11.25 & 11.59 & 11.45 & 10.08 & 17.92 & 27.41 & \underline{38.98} & 9.33 & 18.27 & 11.32 & 16.75 \\
\hline
RankZephyr (7B)
& 19.85 & 19.02 & 12.2 & 33.77 & 24.78 & 13.01 & 23.69 & 29.14 & 32.76 & 6.61 & 24.9 & 30.17 & 22.49 \\
Rank1 (7B) 
& 25.88 & 38.74 & 16.32 & 39.87 & 35.15 & 24.51 & 33.59 & 12.90 & 28.11 & 3.64 & 30.60 & 38.18 & 27.29 \\
Rank-R1 (7B) 
& 21.12 & 27.36 & 18.13 & 36.52 & 30.27 & 10.74 & 29.30 & 16.88 & 10.28 & 3.09 & 14.35 & 30.47 & 20.71 \\
Rearank (7B) 
& 30.46 & 41.35 & 28.55 & 47.08 & 40.72 & 26.10 & 36.49 & 31.25 & 22.24 & 7.76 & 30.17 & 40.52 & 31.89 \\
ReasonRank (7B)
& 34.19 & 46.16 & 30.68 & \textbf{56.71} & \underline{48.90} & 28.65 & 42.30 & 22.90 & 25.74 & 7.42 & 38.56 & 42.36 & 35.38 \\
\hline
Rank-R1 (14B) & 27.42 & 36.75 & 24.81 & 43.02 & 37.18 & 28.22 & 36.73 & 21.24 & 20.58 & 7.11 & 32.09 & 35.58 & 29.23 \\
\hline
Rank1 (32B)
& 25.63 & 32.28 & 17.63 & 42.72 & 35.25 & 23.55 & 31.08 & 12.24 & \textbf{41.10} & 5.67 & 29.64 & 39.17 & 28.00 \\
Rank-K (32B)
& 28.49 & 38.43 & 24.62 & 48.22 & 43.48 & 30.97 & 32.29 & 25.98 & 23.00 & 9.25 & 35.93 & 42.74 & 31.95 \\
ReasonRank (32B)
& 34.49 & 47.16 & 30.97 & 54.65 & \textbf{52.11} & \underline{33.99} & \textbf{45.06} & 27.77 & 19.99 & 7.94 & 38.49 & 44.83 & \underline{36.45} \\
\midrule

\multicolumn{14}{c}{{\small\textbf{\textit{General LLM Rerankers}}}} \\
\midrule

\multicolumn{14}{l}{Qwen2.5-32B-Instruct} \\
\makecell[l]{\hspace{1.2em}with Original Queries}
& 32.27 & 42.20 & 25.32 & 52.18 & 46.71 & 28.82 & 41.99 & 17.82 & 27.89 & 7.16 & 27.59 & 43.02 & 32.75 \\
\makecell[l]{\hspace{1.2em}with GPT-4}
& 30.59 & 45.01 & \underline{31.08} & 52.72 & 41.38 & 28.21 & 37.09 & 22.34 & 23.50 & 6.80 & 35.35 & 41.66 & 32.98 ($\uparrow$0.23)\\
\makecell[l]{\hspace{1.2em}with DeepRetrieval}
& 30.51 & 43.59 & 25.63 & 51.01 & 47.99 & 29.39 & 41.65 & 19.49 & 28.02 & 7.18 & 27.16 & 42.11 & 32.81 ($\uparrow$0.06)\\
\makecell[l]{\hspace{1.2em}with TongSearch (7B)}
& 30.08 & 41.35 & 26.49 & 49.27 & 47.28 & 27.17 & 41.81 & 20.11 & 24.44 & 7.48 & 28.67 & 41.50 & 32.14 ($\downarrow$0.61) \\

\makecell[l]{\hspace{1.2em}with \textbf{QReason}}
& \cellcolor{gray!15}\underline{35.70}
& \cellcolor{gray!15}44.40
& \cellcolor{gray!15}29.44
& \cellcolor{gray!15}51.46
& \cellcolor{gray!15}48.39
& \cellcolor{gray!15}31.11
& \cellcolor{gray!15}42.56
& \cellcolor{gray!15}18.06
& \cellcolor{gray!15}34.06
& \cellcolor{gray!15}\underline{9.66}
& \cellcolor{gray!15}32.51
& \cellcolor{gray!15}\underline{44.86}
& \cellcolor{gray!15}35.18 ($\uparrow$2.43) \\

\addlinespace[2pt]
\multicolumn{14}{l}{Qwen3-30B-A3B-Instruct-2507} \\
\makecell[l]{\hspace{1.2em}with Original Queries}
& 33.23 & \underline{49.21} & 27.54 & 50.39 & 43.90 & \textbf{34.22} & 40.81 & 27.50 & 21.95 & 9.11 & 33.96 & 41.90 & 34.48 \\
\makecell[l]{\hspace{1.2em}with GPT-4}
& 29.33 & 46.33 & 28.09 & 47.72 & 42.94 & 28.03 & 35.80 & 25.86 & 18.29 & 9.47 & 30.80 & 39.82 & 31.87 ($\downarrow$2.61)\\
\makecell[l]{\hspace{1.2em}with DeepRetrieval}
& 32.17 & 48.33 & 30.24 & 51.15 & 46.30 & 32.86 & 42.69 & 29.18 & 20.13 & \textbf{10.02} & 31.78 & 41.50 & 34.70 ($\uparrow$0.22)\\
\makecell[l]{\hspace{1.2em}with TongSearch (7B)}
& 32.88 & 46.26 & 25.96 & 49.38 & 43.21 & 32.73 & 38.87 & 27.48 & 20.22 & 7.73 & 31.63 & 39.90 & 33.02 ($\downarrow$1.46) \\
\makecell[l]{\hspace{1.2em}with \textbf{QReason}}
& \cellcolor{gray!15}\textbf{36.58} & \cellcolor{gray!15}\textbf{49.61} & \cellcolor{gray!15}28.39 & \cellcolor{gray!15}52.83 & \cellcolor{gray!15}45.80 & \cellcolor{gray!15}33.33 & \cellcolor{gray!15}43.98 & \cellcolor{gray!15}28.35 & \cellcolor{gray!15}19.68 & \cellcolor{gray!15}9.09 & \cellcolor{gray!15}35.04 & \cellcolor{gray!15}43.73 & \cellcolor{gray!15}35.53 ($\uparrow$1.05) \\

\addlinespace[2pt]
\multicolumn{14}{l}{Qwen3.5-35B-A3B} \\
\makecell[l]{\hspace{1.2em}with Original Queries}
& 32.07 & 43.89 & \textbf{32.06} & 53.89 & 47.94 & 30.26 & 39.61 & 28.75 & 27.24 & 8.26 & 40.28 & 41.58 & 35.49 \\
\makecell[l]{\hspace{1.2em}with GPT-4}
& 30.35 & 47.83 & 29.95 & 55.74 & 45.27 & 31.18 & 40.36 & 30.26 & 27.56 & 8.43 & 39.65 & 43.15 & 35.81 ($\uparrow$0.32)\\
\makecell[l]{\hspace{1.2em}with DeepRetrieval}
& 32.34 & 45.35 & 30.49 & 53.47 & 46.82 & 29.66 & 40.53 & 27.30 & 24.57 & 8.73 & \underline{41.37} & 43.15 & 35.32 ($\downarrow$0.17) \\
\makecell[l]{\hspace{1.2em}with TongSearch (7B)}
& 33.84 & 44.10 & 30.22 & 53.27 & 45.68 & 29.22 & 39.84 & 30.72 & 28.79 & 8.47 & 40.95 & 43.23 & 35.69 ($\uparrow$0.20) \\
\makecell[l]{\hspace{1.2em}with \textbf{QReason}}
& \cellcolor{gray!15}35.34 & \cellcolor{gray!15}46.63 & \cellcolor{gray!15}30.46 & \cellcolor{gray!15}\underline{56.44} & \cellcolor{gray!15}47.12 & \cellcolor{gray!15}29.52 & \cellcolor{gray!15}\underline{44.05} & \cellcolor{gray!15}\underline{31.55}& \cellcolor{gray!15}25.70 & \cellcolor{gray!15}9.17 & \cellcolor{gray!15}\textbf{41.73} & \cellcolor{gray!15}\textbf{45.49} & \cellcolor{gray!15}\textbf{36.93} ($\uparrow$1.44) \\

\bottomrule
\end{tabular}
\label{tab:bright-main-results}
\end{table*}

\subsection{Training and Inference} 
\paragraph{Training Stage.}
In \textbf{RGST}, we initialize the rewriter from Qwen2.5-7B-Instruct and optimize it using the causal language modeling objective in Eq.~(\ref{eq:L_RGST}). In \textbf{RADR}, we continue reinforcement learning from the RGST-trained rewriter. During rollout, the gold relevant-passages $\mathcal{P}^{+}$ are hidden from the policy model and used only for reward computation. We optimize the rewriter using the clipped GRPO objective in Eq.~(\ref{eq:L_RADR}).
\paragraph{Inference Stage.}
At inference time, the rewriter operates autonomously and generates the reasoning query $q'$ from the original query $q$ alone. Although $\mathcal{P}^{+}$ is unavailable at this stage, it acts as a critical supervisory anchor that makes relevance-grounded reasoning observable during training, allowing the model to internalize the relevance patterns revealed by $(q, \mathcal{P}^{+})$ pairs. In this sense, $\mathcal{P}^{+}$ serves as a supervisory scaffold that is removed after training. The reranker then takes both $q$ and $q'$, with $q'$ directly appended to $q$, and performs non-reasoning ranking.

\section{Experiment}
\subsection{Dataset}
\paragraph{Training Data Construction.}
Our training data is constructed based on the public \texttt{reasonrank\_data\_13k} dataset released by ReasonRank \cite{reasonrank_arxiv}. We use DeepSeek-R1 \cite{deepseek-r1} as the teacher model to construct the RGST training data. 
\paragraph{Evaluation Benchmark.}
We evaluate our method on BRIGHT \cite{bright}, a benchmark for reasoning-intensive retrieval. Since relevant documents often require deeper reasoning beyond lexical or shallow semantic matching, BRIGHT is well suited for evaluating complex reranking. We use NDCG@10 as the evaluation metric. In addition to BRIGHT, we also report supplementary results on FreshStack \cite{freshstack} and NeuCLIRBench \cite{neuclirbench} in Appendix~\ref{app:additional_benchmarks} to further examine the effectiveness of our method.


\subsection{Baselines}


The baselines used in our experiments can be divided into two groups: task-specific rerankers and general LLM rerankers. 
\paragraph{Task-Specific Rerankers.}
We include several models specifically trained or optimized for reranking, including RankT5 (3B) \cite{rankt5}, RankZephyr (7B) \cite{rankzephyr}, Rearank (7B) \cite{rearank}, Rank-R1 (7B and 14B) \cite{rankr1}, Rank1 (7B and 32B) \cite{rank1}, Rank-K (32B) \cite{rankk} and ReasonRank (7B and 32B) \cite{reasonrank_arxiv}. Among them, Rearank (7B), Rank-R1 (7B and 14B), Rank1 (7B and 32B), Rank-K (32B), and ReasonRank (7B and 32B) are reasoning rerankers, while RankT5 (3B) and RankZephyr (7B) are non-reasoning rerankers.
\paragraph{General LLM Rerankers.}
We use Qwen2.5-32B-Instruct \cite{qwen2_5}, Qwen3-30B-A3B-Instruct-2507 \cite{qwen3}, and Qwen3.5-35B-A3B \cite{qwen3_5} as rerankers, all of which perform non-reasoning reranking. On top of these rerankers, we compare multiple rewriters or rewriting methods, including GPT-4 \cite{gpt4}, DeepRetrieval \cite{deepretrieval}, TongSearch (7B) \cite{tongsearch}, and our method. We do not pair rewritten queries with rerankers specifically trained for reranking, because these models are trained on original queries and are therefore adapted to the distribution of original-query inputs.


Following ReasonRank, we use the GPT-4-rewritten queries provided by BRIGHT as retrieval queries, and adopt ReasonIR (8B) \cite{reasonir} as the initial retriever. We then rerank the top-100 passages retrieved by ReasonIR.

\subsection{Main Results}
As shown in Table~\ref{tab:bright-main-results}, QReason achieves the best overall performance on BRIGHT when paired with Qwen3.5-35B-A3B, reaching an average score of 36.93 and outperforming reasoning reranker baselines of a similar scale, including Rank1 (32B), Rank-K (32B), and the strongest baseline, ReasonRank (32B). This result is particularly notable because these baselines require generating reasoning traces multiple times during reranking, whereas QReason only needs to generate the reasoning query once and then performs non-reasoning reranking. The gain over ReasonRank (32B) therefore suggests that decoupling reusable query-level reasoning from window-specific ranking can reduce repeated reasoning generation and improve ranking efficiency, while still achieving performance comparable to, or even better than, the strongest reasoning reranking model.

QReason also delivers the most consistent improvements across different rerankers and is substantially more stable than existing rewriting baselines. Specifically, when applied to Qwen2.5-32B-Instruct, Qwen3-30B-A3B-Instruct-2507, and Qwen3.5-35B-A3B, QReason consistently improves the original-query setting by 2.43, 1.05, and 1.44 points, respectively. In contrast, the other rewriting methods show much less stable behavior. For example, GPT-4 drops by 2.61 points on Qwen3-30B-A3B-Instruct-2507, and TongSearch (7B) reduces performance on both Qwen2.5-32B-Instruct and Qwen3-30B-A3B-Instruct-2507. These results indicate that generic rewriting does not reliably benefit reranking, whereas QReason produces ranking-oriented reasoning queries that transfer more robustly across different rerankers.

\subsection{Ablation Study}


\begin{table}[t]
\centering
\small
\setlength{\tabcolsep}{3pt}
\renewcommand{\arraystretch}{1.12}
\caption{Ablation study of QReason on BRIGHT using Qwen2.5-32B-Instruct as the reranker.}
\label{tab:ablation}
\begin{tabular}{p{0.34\columnwidth}p{0.42\columnwidth}c}
\toprule
Variant & Setting & Avg. \\
\midrule
QReason & Full & \textbf{35.18} \\
\midrule
\multirow{2}{=}{RGST and RADR Training}
& Only RGST & 33.50 \\
& Only RADR & 33.91 \\
\midrule
Relevant-Passages Guidance
& RGST w/o $\mathcal{P}^{+}$ & 32.98 \\
\midrule
\multirow{2}{=}{Reward Design}
& w/o Pointwise Reward & 33.85 \\
& w/o Listwise Reward & 33.90 \\
\bottomrule
\end{tabular}
\end{table}

\begin{table*}[t]
\centering
\fontsize{9.5pt}{11pt}\selectfont
\setlength{\tabcolsep}{6pt}
\renewcommand{\arraystretch}{1.1}
\caption{End-to-end latency comparison on six BRIGHT subsets. For QReason, the total latency is measured end-to-end, including both the one-time query rewriting cost and the downstream reranking cost. Unit: seconds.}
\label{tab:empirical-efficiency}
\begin{tabular}{lccccc}
\toprule
\multirow{2}{*}{Dataset} & ReasonRank (32B) & \multicolumn{2}{c}{\textbf{QReason} + Qwen3.5-35B-A3B} & \multicolumn{2}{c}{\textbf{QReason} + Qwen2.5-32B-Instruct} \\
\cmidrule(lr){2-2} \cmidrule(lr){3-4} \cmidrule(lr){5-6}
& Cost & Cost & Speedup & Cost & Speedup \\
\midrule
economic            & 1056.08 & 163.83 & 6.45$\times$  & 503.97  & 2.10$\times$ \\
biology             & 754.67  & 95.89  & 7.87$\times$  & 277.92  & 2.72$\times$ \\
sustainable\_living & 820.03  & 130.01 & 6.31$\times$  & 368.01  & 2.23$\times$ \\
leetcode            & 3385.11 & 315.04 & 10.75$\times$ & 1193.75 & 2.84$\times$ \\
aops                & 1742.63 & 228.83 & 7.62$\times$ & 663.01  & 2.63$\times$ \\
theoremqa\_theorems & 1188.31 & 205.18 & 5.79$\times$  & 456.79  & 2.60$\times$ \\
\midrule
avg                 & 1491.14 & \textbf{189.80} & \textbf{7.86$\times$}  & \textbf{577.24}  & \textbf{2.58$\times$} \\
\bottomrule
\end{tabular}
\end{table*}

Table~\ref{tab:ablation} reports the ablation results of QReason on BRIGHT using Qwen2.5-32B-Instruct as the reranker.

\paragraph{The Effectiveness of RGST and RADR Training.}
A clear performance drop is observed when either training stage is removed. Using only RGST leads to a decline of 1.68 points in Avg. NDCG@10, while using only RADR results in a drop of 1.27 points. This shows that the two stages play complementary roles. RGST provides stable evidence-grounded initialization for reasoning query generation, while RADR further aligns the rewritten queries with downstream reranking objectives. 
\paragraph{The Effectiveness of Relevant-Passages Guidance in RGST.}
Excluding gold relevant-passages from RGST causes a significant degradation of 2.20 points. This result verifies that introducing $\mathcal{P}^{+}$ during supervised tuning is critical for grounding the rewriting process in explicit relevance evidence. Without such grounding, the generated reasoning queries are more likely to drift away from the key facets required for accurate ranking.
\paragraph{The Effectiveness of Reward Design in RADR.}
Removing either reward term in RADR also degrades performance. Specifically, removing the pointwise reward causes a drop of 1.33 points, while removing the listwise reward results in a decline of 1.28 points. These results indicate that the two rewards are complementary: the listwise reward directly optimizes ranking quality at the whole-list level, whereas the pointwise reward constrains the rewritten query to remain consistent with the gold relevant-passages.

\subsection{Efficiency Analysis}
The total generation cost of conventional reasoning listwise rerankers can be expressed as
\begin{equation}
G_{\mathrm{reason}}=\sum_{j=1}^{K}\bigl(\ell(c_j)+\ell(\pi_j)\bigr)\approx K(R+O),
\end{equation}
where $\ell(\cdot)$ denotes the output length, $K$ is the number of sliding windows, and $R$ and $O$ denote the average lengths of the reasoning trace and ranking output, respectively. Likewise, the total generation cost of QReason is
\begin{equation}
G_{\mathrm{QReason}}=\ell(q')+\sum_{j=1}^{K}\ell(\pi'_j)\approx R'+KO,
\end{equation}
where $R'=\ell(q')$ is the length of the rewritten reasoning query. Compared with the conventional paradigm, QReason replaces repeated window-specific reasoning generation with a one-time rewriting step. In practice, with $R \approx R'$, the efficiency gain mainly comes from avoiding reasoning generation in the remaining $K-1$ windows.

It is worth noting that QReason appends $q'$ to each ranking prompt, which slightly increases the input length of each window. However, this extra cost is introduced in the prefilling stage, whereas conventional reasoning rerankers repeatedly generate full reasoning traces through autoregressive decoding, which is inherently sequential and more latency-sensitive in practice.

\begin{table*}[t]
\centering
\footnotesize
\setlength{\tabcolsep}{4pt}
\renewcommand{\arraystretch}{1.15}
\caption{Case study of QReason and TongSearch rewrites. \textcolor{green!50!black}{Green text} denotes content aligned with the gold evidence, while \textcolor{red}{red text} denotes unsupported details introduced in the rewrite.}
\label{tab:case-study}
\begin{tabularx}{0.98\textwidth}{p{0.20\textwidth}p{0.22\textwidth}X X}
\toprule
Query Intent & Gold Evidence & QReason Rewrite & TongSearch (7B) Rewrite \\
\midrule
partition household food expenditures into BEA producing industries and attribute expenditures through an input-output framework.
&
\textcolor{green!50!black}{input-output accounting framework; supply-use / make-use table system; industry-commodity mapping; expenditure attribution through production chains.}
&
\textbf{Matched:} \textcolor{green!50!black}{input-output accounting framework; supply-use / make-use table system; industry-commodity mapping.} \newline
\textbf{Noise:} \textcolor{red}{national accounting systems; expenditure survey methodology.}
&
\textbf{Matched:} \textcolor{green!50!black}{input-output accounting framework.} \newline
\textbf{Noise:} \textcolor{red}{Food-at-Home / Food-Away-from-Home mapping; direct allocation to 311FT, 445, and 722; assumed expenditure shares; splitting strategy.}
\\
\bottomrule
\end{tabularx}
\end{table*}

We further compare the empirical efficiency of two QReason instantiations against ReasonRank (32B) on six BRIGHT subsets. As shown in Table~\ref{tab:empirical-efficiency}, QReason yields substantial end-to-end efficiency gains over explicit reasoning reranking. After including the rewriting cost, QReason still achieves about $7.9\times$ speedup when using the Mixture-of-Experts (MoE) reranker Qwen3.5-35B-A3B and about $2.6\times$ speedup when using Qwen2.5-32B-Instruct, compared with ReasonRank (32B). These results show that QReason substantially alleviates the latency bottleneck of reasoning listwise rerankers. 

\subsection{Parameter Sensitivity}


We study the weighting coefficient $\lambda$ in RADR, which balances the listwise and pointwise rewards. We vary $\lambda$ from 0.1 to 1.5, train the rewriter under the same setting, and evaluate the resulting rewritten queries using Qwen2.5-32B-Instruct as the reranker. The performance exhibits a clear non-monotonic trend: it first improves as $\lambda$ increases and then declines when $\lambda$ becomes too large, with the best result obtained at $\lambda=0.6$. This indicates that a moderate pointwise reward can regularize the semantic scope of the rewritten query without weakening its downstream reranking utility.

\subsection{Case Study}
\label{subsec:case-study}
Table~\ref{tab:case-study} presents a representative case for qualitatively comparing different rewritten queries. \textbf{Query Intent} summarizes the main information need in the original query, and \textbf{Gold Evidence} lists the core evidence reflected in the relevant passages. Under each rewritten query, \textbf{Matched} refers to content aligned with the gold evidence, whereas \textbf{Noise} refers to additional content not directly supported by the gold relevant-passages. Owing to space constraints, we report only the most salient phrases. Full case details are provided in Appendix~\ref{app:case_details}.

In the case shown in the table, the original query asks how household food expenditures can be attributed to BEA producing industries under an input-output framework for pollution-impact analysis. Accordingly, the gold evidence centers on the accounting structure needed for such attribution, including the input-output framework, the supply-use / make-use table system, and industry-commodity mapping. QReason includes the input-output framework, supply-use tables, and industry-level linkage structure emphasized in the gold evidence, together with some broad additional content such as national accounting systems and expenditure survey methodology. TongSearch also covers part of the same evidence space, but further introduces more specific allocation details, which are not explicitly supported by the gold evidence.

\section{Conclusion}

In this paper, we propose QReason, a decoupled reranking framework that reuses query-level reasoning across sliding windows. Existing reasoning-based rerankers repeatedly generate highly similar and redundant reasoning traces across sliding windows. QReason addresses this issue by training a rewriter to generate a ranking-oriented reasoning query through Relevance-Grounded Supervised Tuning and Rank-Aligned Dual-Reward Refinement. The generated reasoning query is then reused by non-reasoning rerankers across different windows. Experiments on BRIGHT demonstrate that QReason consistently improves different LLM rerankers and achieves strong performance without redundant reasoning generation.

\section*{Limitations}
QReason optimizes reasoning queries with listwise and pointwise relevance rewards, which do not explicitly capture multidimensional requirements such as coverage, faithfulness, and redundancy. Rubrics offer an interpretable way to specify such criteria \citep{liu2026rules}, and have recently been applied to document reranking \citep{liu2026rubricranker}. Future work could incorporate query-specific rubric rewards into RADR to improve the controllability and robustness of reusable reasoning queries.

\section*{Ethical Considerations}

This work focuses on query rewriting and passage reranking for information retrieval. It does not involve the collection of new user data, personal information, or sensitive attributes, and does not include human-subject studies or human annotation. All experiments are conducted on publicly available datasets and pretrained models, and we cite the corresponding sources. The main potential risk is that errors in rewriting or reranking may affect the order of retrieved passages and expose users to less accurate or less reliable information. This risk is common to retrieval and reranking systems, and the proposed method is intended for research use rather than direct deployment in high-stakes scenarios.

\section*{Acknowledgments}
This work was supported by the National Natural Science Foundation of China No. 62472038 and  No. 62437001.


\bibliography{custom}

\appendix

\section{Evaluation Benchmark and Baselines}
\subsection{Evaluation Benchmark}
In our experiments, we use BRIGHT \cite{bright} as the evaluation benchmark. BRIGHT is a reasoning-intensive retrieval benchmark designed to evaluate whether retrieval systems can identify relevant documents beyond lexical or shallow semantic matching. It contains 1,384 real-world queries from diverse domains, including economics, psychology, mathematics, and coding, with queries derived from naturally occurring data and carefully curated human inputs. Unlike conventional retrieval benchmarks where surface-form or semantic similarity is often sufficient, BRIGHT requires models to infer implicit connections between the query and candidate documents. For example, resolving a coding query may require understanding the underlying program logic, while mathematical and scientific queries may require matching the query to relevant concepts, theorems, or explanatory evidence. Therefore, BRIGHT provides a challenging testbed for evaluating retrieval and reranking methods under reasoning-intensive scenarios.
\subsection{Baselines} 
\paragraph{ReasonIR.}
ReasonIR-8B \cite{reasonir} is a LLaMA3.1-8B-based bi-encoder retriever specifically trained for reasoning-intensive retrieval, using synthetic reasoning-oriented data with challenging queries and hard negatives.
\paragraph{RankT5.}
RankT5 \cite{rankt5} is a T5-based pointwise reranker optimized with a ranking loss function, which scores each query-document pair independently for relevance estimation.
\paragraph{RankZephyr.}
RankZephyr \cite{rankzephyr} is an open-source listwise LLM reranker distilled from GPT-3.5- and GPT-4-generated ranking data, and performs reranking by directly generating an ordered list of candidate passages.
\paragraph{Rearank.}
REARANK \cite{rearank} is a Qwen2.5-7B-based listwise reasoning reranker trained with reinforcement learning, which explicitly generates reasoning before producing the final ranked list of candidate passages.
\paragraph{Rank1.}
Rank1 \cite{rank1} is a pointwise reasoning reranker distilled from DeepSeek-R1 reasoning traces, which estimates relevance by generating a reasoning chain before making a true/false relevance judgment.
\paragraph{Rank-R1.}
Rank-R1 \cite{rankr1} is a setwise reasoning reranker trained with GRPO, which selects the most relevant passage from a candidate set and applies heap sort to construct the final ranking.
\paragraph{Rank-K.}
Rank-K \cite{rankk} is a QwQ-32B-based listwise reasoning reranker distilled from DeepSeek-R1, which generates reasoning traces and applies a sliding-window strategy during inference.
\paragraph{ReasonRank.}
ReasonRank \cite{reasonrank_arxiv} is trained through a two-stage framework on high-quality reasoning-intensive data synthesized by DeepSeek-R1, consisting of a cold-start supervised fine-tuning (SFT) stage and a reinforcement learning (RL) stage. In the RL stage, it is optimized with a multi-view ranking reward. ReasonRank performs listwise reranking with a sliding-window strategy and achieves strong performance across multiple benchmarks.
\paragraph{DeepRetrieval.}
DeepRetrieval \cite{deepretrieval} is an RL-based query generation method that trains LLMs through trial and error without supervised reference queries, using retrieval metrics as rewards to generate queries that improve retrieval performance.
\paragraph{TongSearch.}
TongSearch QR \cite{tongsearch} is a family of small-scale language models for query reasoning and rewriting in reasoning-intensive retrieval, trained with reinforcement learning and a semi-rule-based reward function.

\section{Implementation Details}
\label{sec:appendix}
\subsection{Training Data Construction}
\label{app:training_data_construction}
Our training data is constructed based on the public \texttt{reasonrank\_data\_13k}\footnote{\url{https://huggingface.co/datasets/liuwenhan/reasonrank_data_13k}} dataset released by ReasonRank \cite{reasonrank_arxiv}. The \texttt{reasonrank\_data\_13k} dataset contains 13{,}549 training instances spanning ten source datasets: \texttt{math-theorem}, \texttt{sustainable\_living}, \texttt{msmarco}, \texttt{biology}, \texttt{robotics}, \texttt{math-qa}, \texttt{leetcode}, \texttt{earth\_science}, \texttt{stackoverflow}, and \texttt{economics}. We randomly sample 6{,}000 instances from the full dataset to construct the RGST data, and use the remaining 7{,}549 instances to build the RADR data.

\subsection{Training Details}
We provide the implementation details of the two training stages in QReason, including the training framework, hardware setup, key hyperparameters, data construction, and prompt templates.
\paragraph{RGST.}
In RGST, each training sample is constructed as an input-output pair, where the input consists of the original query $q$ and its gold relevant-passages $\mathcal{P}^{+}$, and the target output is the teacher-synthesized reasoning query $q'$. We use the \texttt{ms-swift}\footnote{\url{https://github.com/modelscope/ms-swift}} framework to fine-tune Qwen2.5-7B-Instruct\footnote{\url{https://huggingface.co/Qwen/Qwen2.5-7B-Instruct}}~\cite{qwen2_5} on 8 A800 GPUs, and the whole stage takes approximately 2 hours. Training is conducted for 3 epochs with a per-device batch size of 2 and 4 gradient accumulation steps. The learning rate is set to $5\times10^{-6}$. To improve training efficiency and memory scalability, we adopt DeepSpeed ZeRO-3 \cite{deepspeedzero} and FlashAttention \cite{flashattn}. 

\begin{figure}[t]\centering
  \includegraphics[width=\columnwidth]{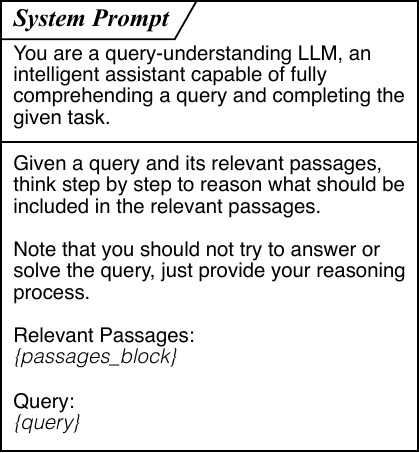}
  \caption{RGST prompt of data synthesis and training.}
  \label{fig:sft_prompt}
\end{figure}

The training data are synthesized from \texttt{reasonrank\_data\_13k} using DeepSeek-R1. For each sample, the teacher model takes the original query together with its gold relevant-passages as input and generates a reasoning query as the supervision target. The same prompt template is used for both data synthesis and RGST training, as shown in Figure~\ref{fig:sft_prompt}.

\begin{figure}[t]
  \includegraphics[width=\columnwidth]{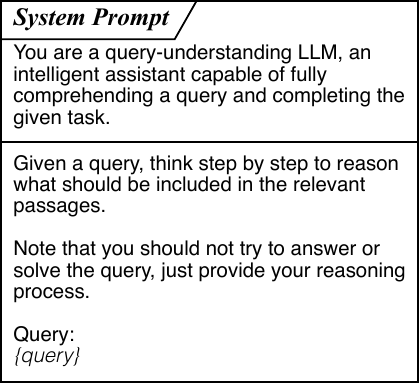}
  \caption{RADR training and QReason inference rewriting prompt template.}
  \label{fig:rewrite_prompt}
\end{figure}

\begin{figure}[t]
  \includegraphics[width=\columnwidth]{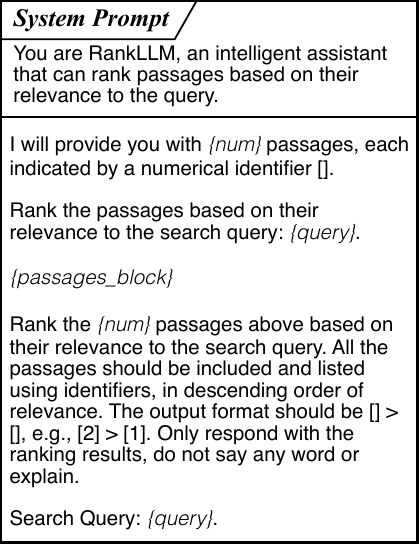}
  \caption{RADR LLM reranker and QReason inference reranking prompt template.}
  \label{fig:rerank_prompt}
\end{figure}

\begin{figure}[t]
  \includegraphics[width=\columnwidth]{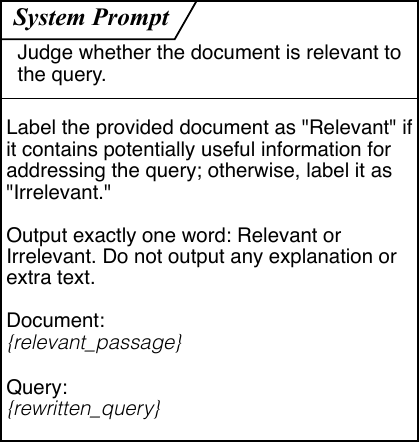}
  \caption{RADR relevance judge model prompt template.}
  \label{fig:rel_judge_prompt}
\end{figure}

\begin{figure}[t]
  \includegraphics[width=\columnwidth]{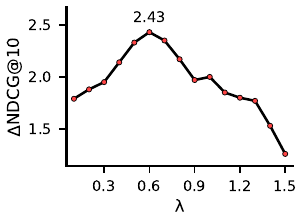}
  \caption{Effect of the reward weight $\lambda$.}
  \label{fig:lambda_delta_ndcg10}
\end{figure}

\paragraph{RADR.}
In RADR, we continue reinforcement learning from the RGST-trained rewriter. During rollout, the rewriter receives only the original query $q$ and autoregressively samples a candidate group $\mathcal{G}(q)$, while the gold relevant-passages $\mathcal{P}^{+}$ are hidden from the policy and used only for reward computation. We further train the RGST checkpoint with GRPO under the \texttt{ms-swift} framework on 8 A800 GPUs, and this stage takes approximately 7 hours. During training, a per-device batch size of 1 is used together with 8 gradient accumulation steps, while 8 sampled generations are drawn for each query to estimate the group-relative advantage. The learning rate is set to $1\times10^{-6}$, the KL coefficient $\beta$ is set to 0.001, and the sampling temperature is 0.8. DeepSpeed ZeRO-3 is also adopted in this stage. The two models used to compute the listwise and pointwise rewards are deployed on a separate server with 2 H100 GPUs and communicate with the training process through a \texttt{vLLM}\footnote{\url{https://github.com/vllm-project/vllm}} server. Specifically, we use Qwen3-30B-A3B as the fixed LLM reranker for the listwise reward and Qwen3-8B as the LLM-based relevance judge for the pointwise reward.

The RADR training data are also constructed from \texttt{reasonrank\_data\_13k}. To align the policy input with the inference setting, the rewriter only observes the original query during generation, while the gold relevant-passages are hidden. The corresponding prompt template for query rewriting is shown in Figure~\ref{fig:rewrite_prompt}. The prompt templates for listwise reranking and pointwise relevance judgment are shown in Figure~\ref{fig:rerank_prompt} and Figure~\ref{fig:rel_judge_prompt}, respectively.

\subsection{Experimental Setup and Inference Details}
For inference, both the rewriter and the reranker are deployed with the \texttt{vLLM} framework, and all experiments are conducted on 4 A800 GPUs. During inference, the rewriter uses the same prompt template as that adopted for rewriting in the RADR stage, as shown in Figure~\ref{fig:rewrite_prompt}. The reranker follows the same prompt template as the LLM reranker used in the listwise reward of RADR, as shown in Figure~\ref{fig:rerank_prompt}. In all experiments, we concatenate the rewritten query after the original query and use the resulting query for reranking.

For the experiments reported in Table~\ref{tab:bright-main-results}, we use a unified evaluation environment to ensure fair comparison. For General LLM Rerankers, we perform listwise reranking with a window size of 20 and a step size of 10. For reranker baselines, we follow the configurations reported in the corresponding papers and reproduce all methods under the same environment. For query rewriting methods, we directly use the reasoning queries generated by TongSearch (7B) and released in its official GitHub repository\footnote{\url{https://github.com/bigai-nlco/TongSearch-QR}}. For DeepRetrieval, we follow its official GitHub\footnote{\url{https://github.com/pat-jj/DeepRetrieval}} implementation and locally reproduce the query rewriting process, while GPT-4 rewrites are taken from the rewritten queries provided in the BRIGHT test set.

The reranking configurations used in Tables~\ref{tab:ablation} and~\ref{tab:empirical-efficiency} are consistent with those in Table~\ref{tab:bright-main-results}. In particular, for the efficiency analysis in Table~\ref{tab:empirical-efficiency}, ReasonRank (32B) is evaluated with the same window size and step size as our method.

In all experiments, we set the decoding temperature of both the rewriter in QReason and the reranker to 0 and use greedy decoding. Therefore, the results are nearly identical across multiple runs.

\section{Additional Experimental Details}
\subsection{Parameter Sensitivity}
As shown in Figure~\ref{fig:lambda_delta_ndcg10}, where the y-axis denotes the NDCG@10 gain of reranking with rewritten queries over original queries, performance first improves and then declines as $\lambda$ increases, with the best result at $\lambda=0.6$. 

When $\lambda$ is small, the pointwise reward contributes little to the overall optimization, causing the listwise reward to dominate. As a result, the rewriter is more likely to overfit to the preference of the fixed reranker, which may lead to semantic drift in the generated reasoning query $q'$.  In contrast, when $\lambda$ becomes too large, the stronger pointwise reward weakens the contribution of the listwise reward, making the rewritten queries less aligned with the reranking objective. 

\subsection{Empirical Efficiency Analysis}
The latency comparison on the full set of datasets is reported in Table~\ref{tab:full-empirical-efficiency}, which also presents the latency of the rewriter in QReason. The average latency of the QReason rewriter is only about $1/65$ of that of ReasonRank (32B) and accounts for approximately $3.9\%$ of the total latency of QReason with Qwen2.5-32B-Instruct. This suggests that the overall runtime of QReason is still dominated by reranking rather than rewriting, making the additional rewriting step highly cost-effective in practice.

\section{More Experiments}
In this section, we provide additional experiments to further evaluate the effectiveness and generalizability of QReason under broader settings, complementing the main results presented in the paper. All results in this section are evaluated using NDCG@10.

\subsection{Extended Main Results on BRIGHT}
As shown in Table~\ref{tab:additional_rerankers_exp}, we further evaluate QReason with additional general LLM rerankers to examine whether its benefit transfers across different downstream models. Specifically, we instantiate the downstream reranker with Qwen2.5-72B-Instruct\footnote{\url{https://huggingface.co/Qwen/Qwen2.5-72B-Instruct}}~\cite{qwen2_5}, Llama3.3-70B-Instruct\footnote{\url{https://huggingface.co/meta-llama/Llama-3.3-70B-Instruct}}~\cite{llama3}, Qwen3-32B\footnote{\url{https://huggingface.co/Qwen/Qwen3-32B}}~\cite{qwen3}, Qwen3-30B-A3B\footnote{\url{https://huggingface.co/Qwen/Qwen3-30B-A3B}}, and Qwen3.6-35B-A3B\footnote{\url{https://huggingface.co/Qwen/Qwen3.6-35B-A3B}}~\cite{qwen3_6}, and compare reranking with original queries against reranking with QReason-generated reasoning queries under the same setting. The results demonstrate that the reasoning queries produced by QReason generalize well across different models and consistently yield clear performance gains.

\subsection{Performance of QReason with Smaller General LLM Rerankers}
We observe that smaller general LLM rerankers show limited ranking effectiveness under the main experimental setting, which may be attributed to their lack of task-specific reranking training. To better examine whether QReason can benefit these smaller rerankers, we conduct additional experiments with a reduced sliding-window configuration. Specifically, we instantiate the downstream reranker with Qwen2.5-7B-Instruct\footnote{\url{https://huggingface.co/Qwen/Qwen2.5-7B-Instruct}}~\cite{qwen2_5}, Llama-3.1-8B-Instruct\footnote{\url{https://huggingface.co/meta-llama/Llama-3.1-8B-Instruct}}~\cite{llama3}, Qwen3-8B\footnote{\url{https://huggingface.co/Qwen/Qwen3-8B}}~\cite{qwen3}, Qwen3.5-9B\footnote{\url{https://huggingface.co/Qwen/Qwen3.5-9B}}~\cite{qwen3_5}, Qwen2.5-14B-Instruct\footnote{\url{https://huggingface.co/Qwen/Qwen2.5-14B-Instruct}}, and Qwen3-14B\footnote{\url{https://huggingface.co/Qwen/Qwen3-14B}}. For each reranker, we compare reranking with original queries against reranking with QReason-generated reasoning queries. We set the window size to 10 and the step size to 5, while keeping all other settings the same as those used in Table~\ref{tab:bright-main-results}. The results are shown in Table~\ref{tab:small_llm_rerankers_bright}.




\subsection{Results on Additional Benchmarks}
\label{app:additional_benchmarks}

We further evaluate QReason on two additional benchmarks, FreshStack \cite{freshstack} and NeuCLIRBench \cite{neuclirbench}, to examine its generalization beyond BRIGHT.

\paragraph{FreshStack.}
FreshStack is a benchmark suite for technical-document retrieval constructed by a reusable IR/RAG benchmark construction framework. It is built from community-asked Stack Overflow questions and technical corpora collected from public GitHub repositories, and covers five fast-growing domains: LangChain, Yolo v7/v8, Laravel 10/11, Angular 16/17/18, and Godot4. It provides nugget-level support judgments for document chunks, making it suitable for evaluating retrieval systems on long, code-intensive technical queries. We use Qwen3-Embedding-8B \cite{qwen3embedding} as the first-stage retriever and rerank the top-100 retrieved passages.

As shown in Table~\ref{tab:freshstack_results}, QReason consistently improves all three LLM rerankers over their original-query settings, with gains of 2.23, 2.53, and 2.26 points on Qwen2.5-32B-Instruct, Qwen3-30B-A3B-Instruct-2507, and Qwen3.5-35B-A3B, respectively. With Qwen3.5-35B-A3B, QReason achieves an average NDCG@10 of 33.63, only 0.05 points below ReasonRank (32B). This result is notable because ReasonRank repeatedly generates reasoning traces during reranking, whereas QReason generates the reasoning query only once and subsequently performs non-reasoning reranking.

\paragraph{NeuCLIRBench.}
NeuCLIRBench is a multilingual retrieval benchmark that supports monolingual, cross-language, and multilingual retrieval. It contains documents originally written in Chinese, Persian, and Russian, together with English machine translations, and combines topics from the TREC NeuCLIR tracks. The monolingual setting evaluates retrieval in English, Persian, Russian, and Chinese; the cross-language setting uses English queries to retrieve documents in each of the three non-English languages; and the multilingual setting retrieves documents across all three languages with English queries. 

Table~\ref{tab:neuclir_results} shows that QReason also transfers consistently to these more traditional multilingual information-retrieval settings. It improves the average performance of all three evaluated rerankers, from 29.73 to 31.13 ($+1.40$) with Qwen2.5-32B-Instruct, from 30.24 to 31.11 ($+0.87$) with Qwen3-30B-A3B-Instruct-2507, and from 30.68 to 31.61 ($+0.93$) with Qwen3.5-35B-A3B. The best configuration, QReason with Qwen3.5-35B-A3B, achieves an average NDCG@10 of 31.61, outperforming ReasonRank (32B) by 1.01 points. Although the gains are smaller than those on BRIGHT and FreshStack, their consistency across all three rerankers indicates that ranking-oriented query reasoning remains beneficial even for less reasoning-intensive monolingual and cross-lingual retrieval tasks.


\section{Case Details}
\label{app:case_details}
\begin{figure*}[t]
    \centering
    \includegraphics[width=\textwidth]{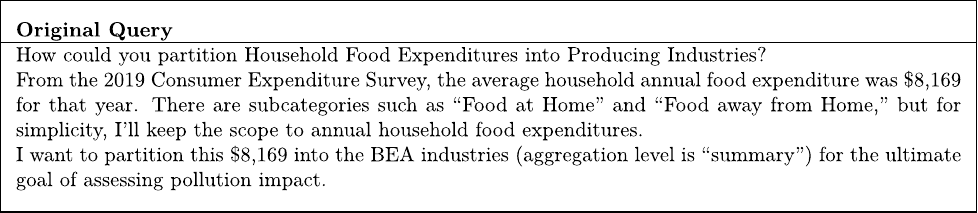}
    \caption{Complete original queries for the case in Section~\ref{subsec:case-study}.}
    \label{fig:case_detail_ori_query}
\end{figure*}

\begin{figure*}[t]
    \centering
    \includegraphics[width=\textwidth]{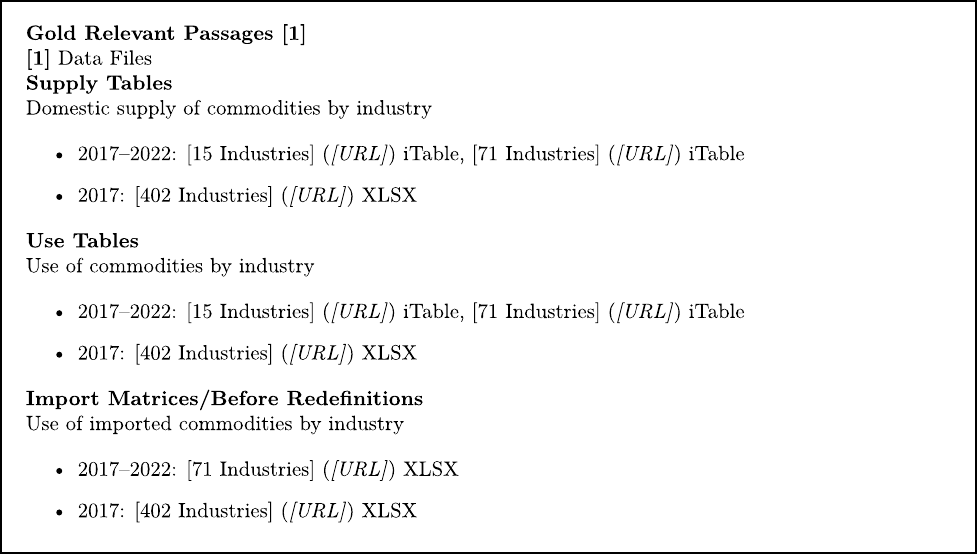}
    \caption{Gold relevant-passages [1] for the case in Section~\ref{subsec:case-study}.}
    \label{fig:case_details_1}
\end{figure*}

\begin{figure*}[t]
    \centering
    \includegraphics[width=\textwidth]{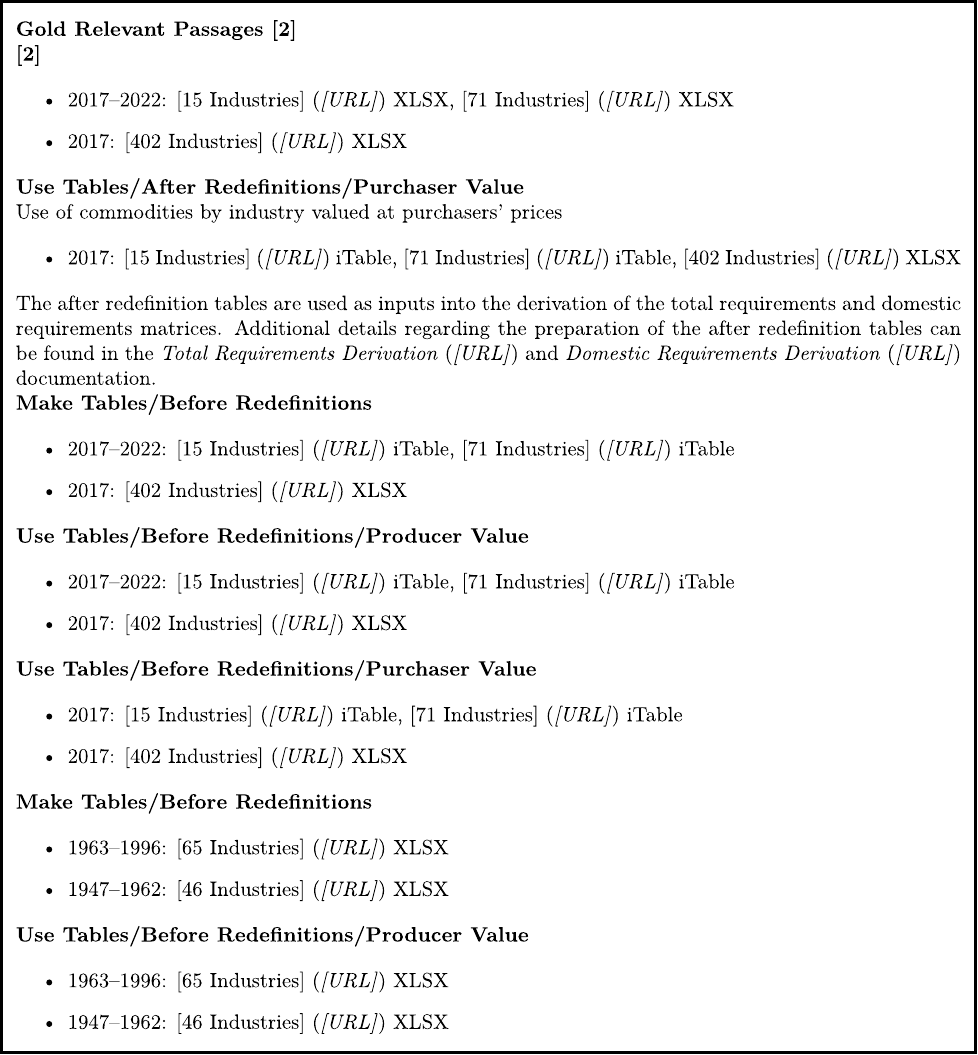}
    \caption{Gold relevant-passages [2] for the case in Section~\ref{subsec:case-study}.}
    \label{fig:case_details_2}
\end{figure*}

\begin{figure*}[t]
    \centering
    \includegraphics[width=\textwidth]{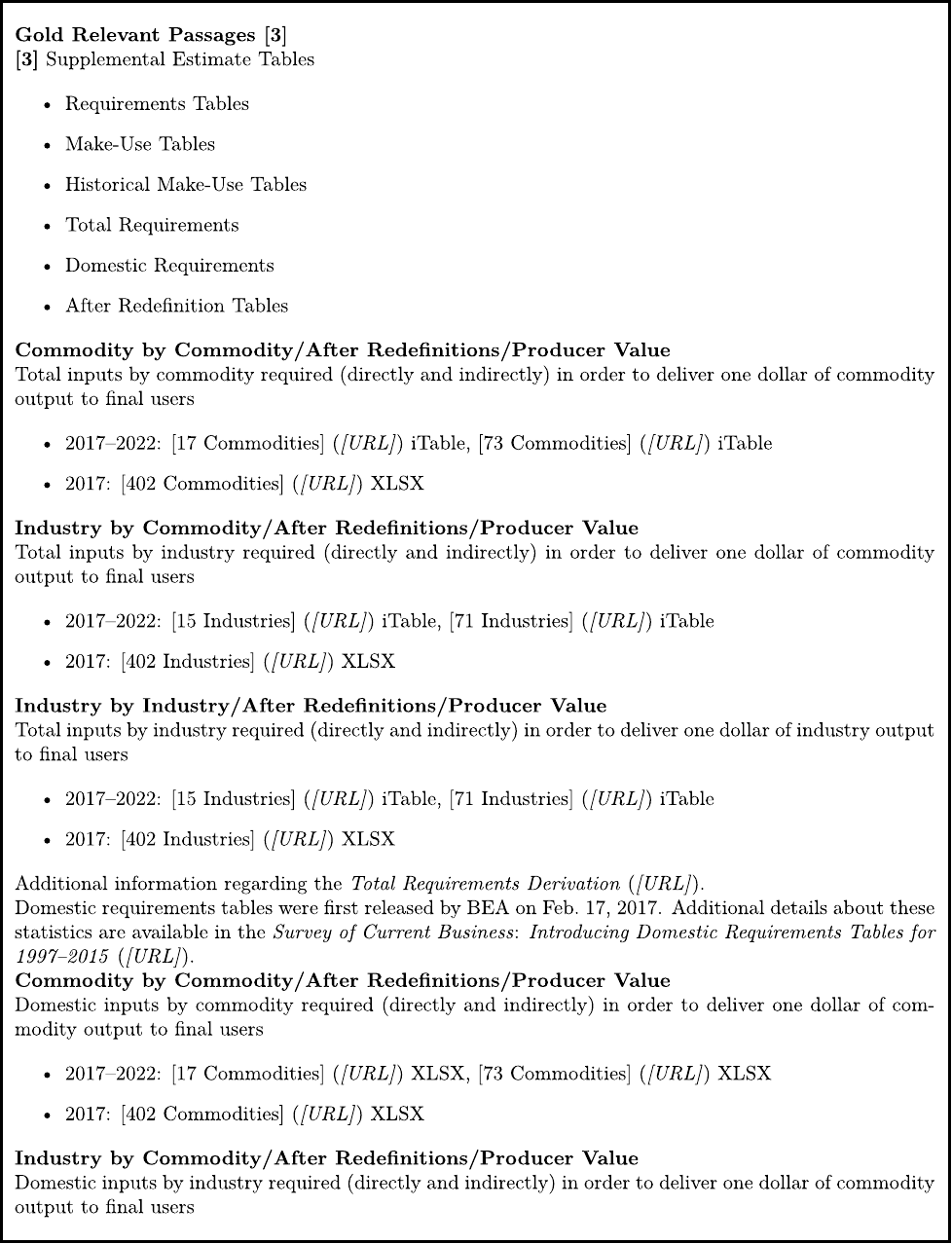}
    \caption{Gold relevant-passages [3] for the case in Section~\ref{subsec:case-study}.}
    \label{fig:case_details_3}
\end{figure*}

\begin{figure*}[t]
    \centering
    \includegraphics[width=\textwidth]{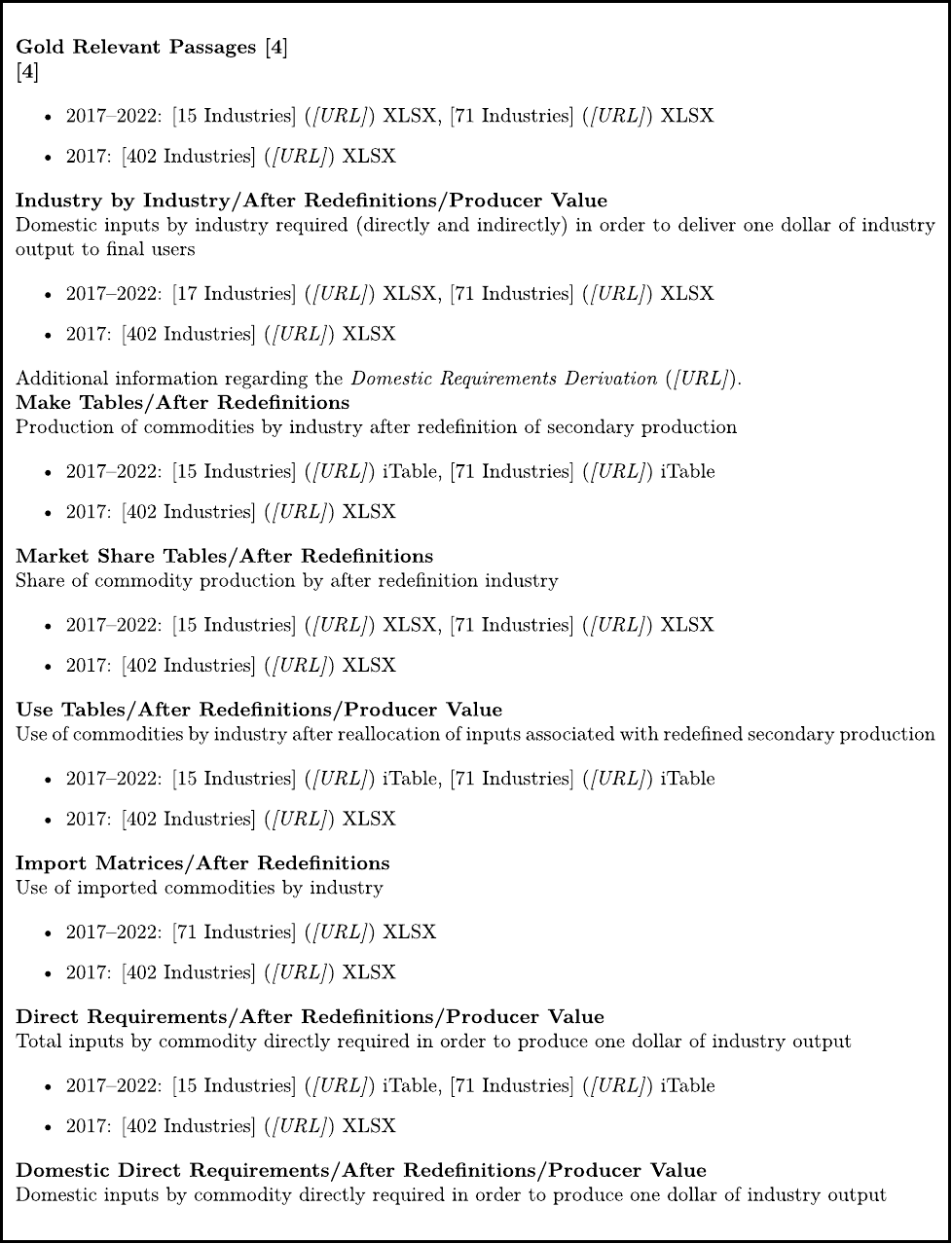}
    \caption{Gold relevant-passages [4] for the case in Section~\ref{subsec:case-study}.}
    \label{fig:case_details_4}
\end{figure*}

\begin{figure*}[t]
    \centering
    \includegraphics[width=\textwidth]{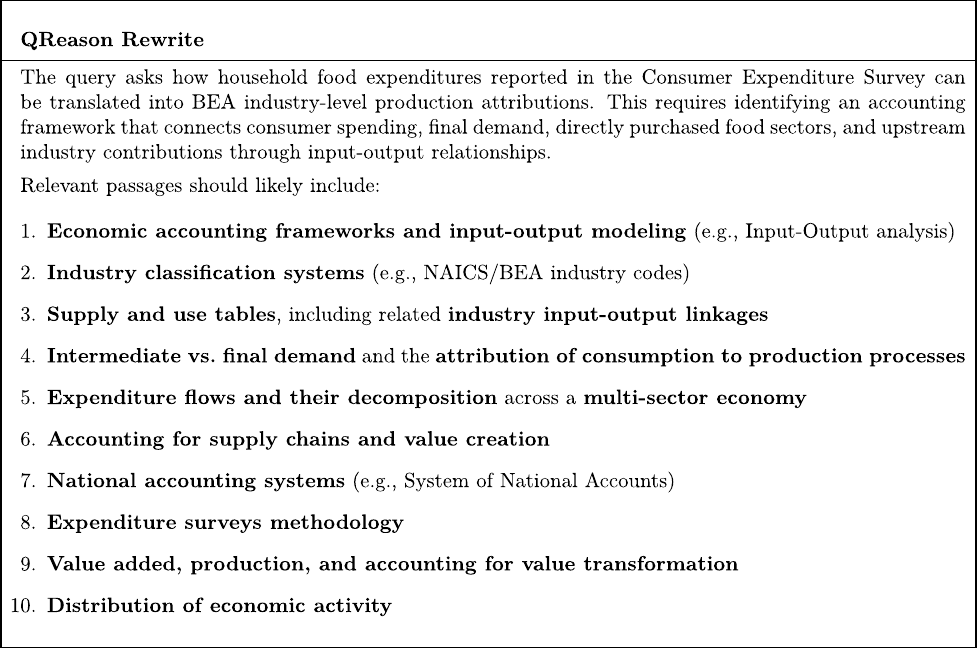}
    \caption{Queries rewritten by QReason for the case in Section~\ref{subsec:case-study}.}
    \label{fig:case_detail_qreason_rewrite}
\end{figure*}

\begin{figure*}[t]
    \centering
    \includegraphics[width=\textwidth]{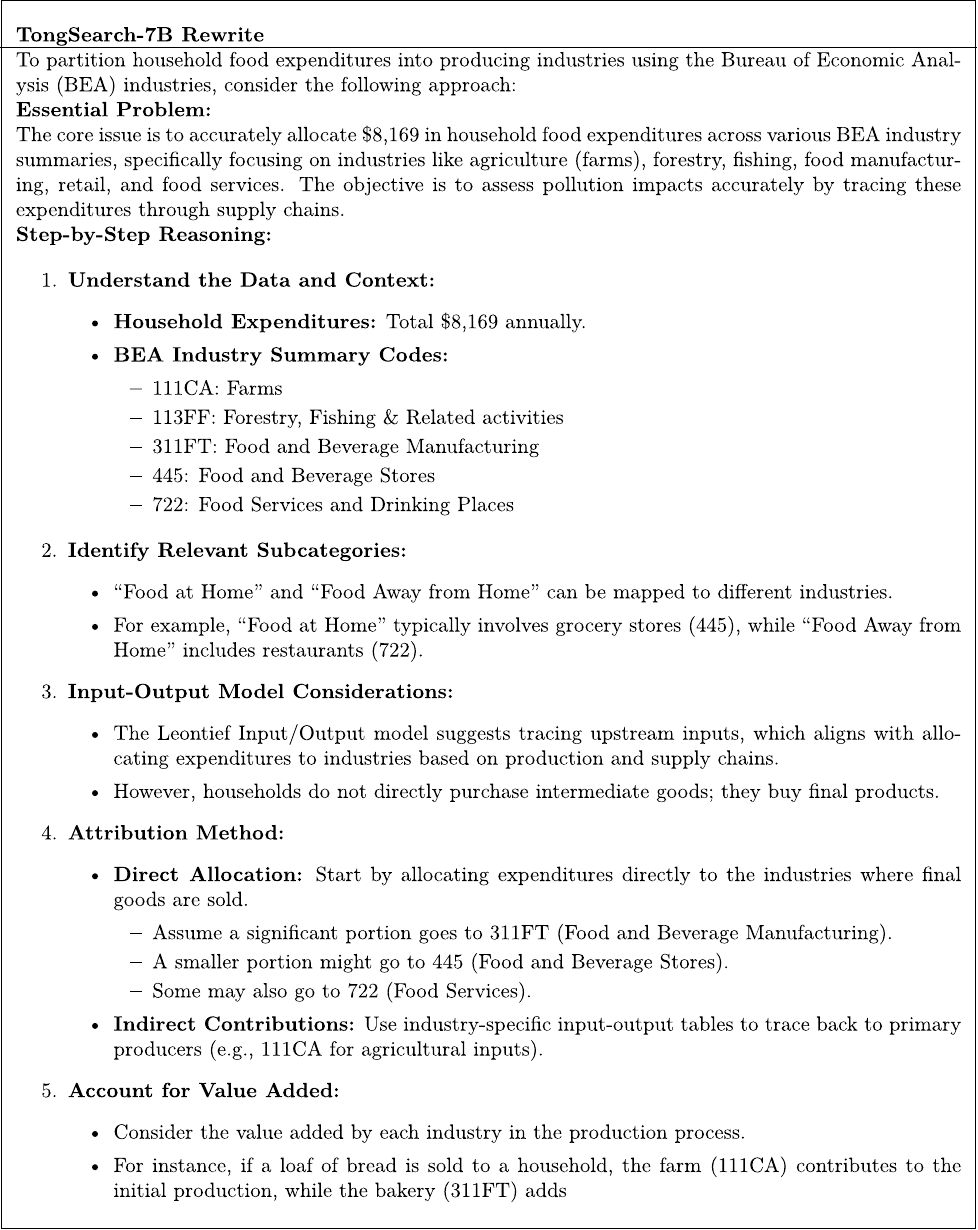}
    \caption{Queries rewritten by TongSearch (7B) for the case in Section~\ref{subsec:case-study}.}
    \label{fig:case_detail_tongsearch_rewrite}
\end{figure*}

We provide the detailed information for the case presented in Section~\ref{subsec:case-study}, including the complete original query, the gold relevant-passages, the query rewritten by QReason, and the query rewritten by TongSearch (7B), as shown in Figures~\ref{fig:case_detail_ori_query}--\ref{fig:case_detail_tongsearch_rewrite}. For readability, URLs in the relevant passages are replaced with \textit{[URL]}.

\section{Use of AI Assistants}
We use ChatGPT to polish the writing of the paper.\footnote{\url{https://chatgpt.com/}}

\section{Artifact Licenses and Intended Use}
All datasets, benchmarks, pretrained models, and software packages used in this work are publicly available. We use them for research purposes and follow their corresponding licenses and terms of use. The datasets and benchmarks, including BRIGHT, FreshStack and \texttt{reasonrank\_data\_13k}, are used under the terms specified by their original providers. The pretrained models, including Qwen, Llama, and DeepSeek models, are used according to their corresponding model licenses or terms of use. The software packages used in our experiments, including \texttt{vLLM} and \texttt{ms-swift}, are used under their respective open-source licenses.

We use existing artifacts only for research purposes, including model training, reranking evaluation, and reproducibility analysis, which is consistent with their intended research or benchmarking use where specified. The artifacts created in this work, including the trained QReason rewriter and generated reasoning queries, are intended for research on query rewriting and passage reranking.

\begin{table*}[t]
\centering
\small
\setlength{\tabcolsep}{4pt}
\renewcommand{\arraystretch}{1.1}
\caption{Full end-to-end latency comparison on all BRIGHT subsets. For QReason, the latency includes both the one-time query rewriting cost and the downstream reranking cost. The rewriter latency is also reported separately. Unit: seconds.}
\label{tab:full-empirical-efficiency}
\begin{tabular}{lcccc}
\toprule
Dataset 
& ReasonRank (32B) 
& QReason + Qwen3.5-35B-A3B 
& QReason + Qwen2.5-32B-Instruct 
& Rewriter \\
\midrule
economic             & 1056.08 & 163.83 & 503.97  & 21.23 \\
earth\_science       & 873.16  & 181.45 & 369.29  & 18.98 \\
robotics              & 1032.78 & 156.06 & 434.76  & 16.75 \\
biology              & 754.67  & 95.89  & 277.92  & 18.27 \\
psychology           & 1121.48 & 180.53 & 494.65  & 18.90 \\
stackoverflow        & 2154.46 & 222.97 & 817.17  & 15.51 \\
sustainable\_living  & 820.03  & 130.01 & 368.01  & 17.71 \\
leetcode             & 3385.11 & 315.04 & 1193.75 & 28.60 \\
pony                 & 943.19  & 116.84 & 349.46  & 22.59 \\
aops                 & 1742.63 & 228.83 & 663.01  & 42.78 \\
theoremqa\_questions & 3437.99 & 552.65 & 1313.49 & 37.44 \\
theoremqa\_theorems  & 1188.31 & 205.18 & 456.79  & 26.33 \\
\midrule
avg                  & 1542.49 & 212.44 & 603.52  & 23.76 \\
\bottomrule
\end{tabular}
\end{table*}

\begin{table*}[t]
\centering
\scriptsize
\setlength{\tabcolsep}{4pt}
\renewcommand{\arraystretch}{1.12}
\caption{Additional results with general LLM rerankers on BRIGHT. Each pair compares reranking with original queries and QReason-rewritten queries under the same reranker.}
\begin{tabular}{lccccccccccccc}
\toprule
Reranker & Econ. & Earth. & Rob. & Bio. & Psy. & Stack. & Sus. & Leet. & Pony & AoPS & TheoQ. & TheoT. & Avg \\
\midrule

\multicolumn{14}{l}{\textbf{Qwen2.5-72B-Instruct}} \\
\makecell[l]{\hspace{1.2em}with Original Queries}
& 29.70 & 41.31 & 24.70 & 47.22 & 45.15 & 28.66 & 38.89 & 22.43 & 21.56 & 5.34 & 31.50 & 43.52 & 31.67 \\
\makecell[l]{\hspace{1.2em}with \textbf{QReason}}
& 33.17 & 43.56 & 29.88 & 50.00 & 46.81 & 29.00 & 43.75 & 23.41 & 30.41 & 7.31 & 36.35 & 45.32 & 34.91 ($\uparrow$3.24) \\
\midrule

\multicolumn{14}{l}{\textbf{Llama3.3-70B-Instruct}} \\
\makecell[l]{\hspace{1.2em}with Original Queries}
& 29.74 & 43.26 & 23.70 & 49.95 & 45.90 & 27.55 & 39.86 & 22.53 & 22.75 & 7.27 & 31.45 & 36.88 & 31.74 \\
\makecell[l]{\hspace{1.2em}with \textbf{QReason}}
& 30.13 & 45.22 & 27.34 & 50.02 & 42.74 & 27.91 & 42.01 & 25.17 & 29.88 & 8.40 & 35.77 & 41.85 & 33.87 ($\uparrow$2.13) \\
\midrule

\multicolumn{14}{l}{\textbf{Qwen3-32B}} \\
\makecell[l]{\hspace{1.2em}with Original Queries}
& 31.83 & 43.34 & 28.12 & 49.02 & 46.98 & 27.33 & 42.43 & 28.31 & 14.80 & 8.05 & 38.12 & 42.39 & 33.39 \\
\makecell[l]{\hspace{1.2em}with \textbf{QReason}}
& 32.99 & 47.39 & 29.59 & 50.33 & 45.25 & 31.51 & 44.69 & 26.95 & 23.72 & 8.28 & 38.78 & 44.94 & 35.37 ($\uparrow$1.98) \\
\midrule

\multicolumn{14}{l}{\textbf{Qwen3-30B-A3B}} \\
\makecell[l]{\hspace{1.2em}with Original Queries}
& 31.82 & 44.91 & 23.31 & 52.77 & 44.62 & 32.01 & 38.00 & 30.10 & 20.89 & 8.77 & 36.81 & 38.61 & 33.55 \\
\makecell[l]{\hspace{1.2em}with \textbf{QReason}}
& 31.74 & 48.23 & 25.27 & 53.59 & 45.56 & 33.44 & 38.61 & 29.46 & 24.64 & 9.18 & 38.29 & 44.36 & 35.20 ($\uparrow$1.65) \\
\midrule

\multicolumn{14}{l}{\textbf{Qwen3.6-35B-A3B}} \\
\makecell[l]{\hspace{1.2em}with Original Queries}
& 33.08 & 45.94 & 28.55 & 51.89 & 48.11 & 28.18 & 39.97 & 30.65 & 29.85 & 9.16 & 40.44 & 40.55 & 35.53 \\
\makecell[l]{\hspace{1.2em}with \textbf{QReason}}
& 34.59 & 47.50 & 29.27 & 54.20 & 46.14 & 30.45 & 42.94 & 29.99 & 31.72 & 10.18 & 43.89 & 44.83 & 37.14 ($\uparrow$1.61) \\
\bottomrule
\end{tabular}
\label{tab:additional_rerankers_exp}
\end{table*}

\begin{table*}[t]
\centering
\scriptsize
\setlength{\tabcolsep}{4pt}
\renewcommand{\arraystretch}{1.12}
\caption{Performance of smaller general LLM rerankers on BRIGHT under a window size of 10 and a step size of 5.}
\begin{tabular}{lccccccccccccc}
\toprule
Reranker & Econ. & Earth. & Rob. & Bio. & Psy. & Stack. & Sus. & Leet. & Pony & AoPS & TheoQ. & TheoT. & Avg \\
\midrule

\multicolumn{14}{l}{\textbf{Qwen2.5-7B-Instruct}} \\
\makecell[l]{\hspace{1.2em}with Original Queries}
& 28.32 & 40.78 & 20.64 & 42.12 & 39.78 & 26.89 & 30.12 & 30.27 & 19.40 & 4.23 & 28.97 & 37.09 & 29.05 \\
\makecell[l]{\hspace{1.2em}with \textbf{QReason}}
& 27.69 & 42.63 & 23.21 & 44.62 & 39.43 & 26.97 & 31.71 & 31.05 & 24.38 & 6.14 & 32.47 & 38.89 & 30.77 ($\uparrow$1.72) \\
\midrule

\multicolumn{14}{l}{\textbf{Llama-3.1-8B-Instruct}} \\
\makecell[l]{\hspace{1.2em}with Original Queries}
& 23.97 & 36.04 & 20.78 & 43.89 & 36.37 & 21.22 & 27.75 & 29.82 & 18.35 & 6.59 & 23.81 & 27.02 & 26.30 \\
\makecell[l]{\hspace{1.2em}with \textbf{QReason}}
& 24.13 & 36.43 & 22.38 & 48.82 & 38.43 & 27.57 & 30.55 & 27.28 & 21.03 & 7.49 & 26.98 & 28.03 & 28.26 ($\uparrow$1.96) \\
\midrule

\multicolumn{14}{l}{\textbf{Qwen3-8B}} \\
\makecell[l]{\hspace{1.2em}with Original Queries}
& 29.53 & 43.00 & 25.03 & 49.47 & 38.22 & 26.37 & 33.71 & 30.48 & 25.08 & 7.68 & 34.49 & 40.20 & 31.94 \\
\makecell[l]{\hspace{1.2em}with \textbf{QReason}}
& 29.07 & 46.23 & 25.42 & 52.44 & 40.51 & 27.28 & 35.58 & 30.61 & 35.18 & 9.41 & 36.74 & 42.55 & 34.25 ($\uparrow$2.31) \\
\midrule

\multicolumn{14}{l}{\textbf{Qwen3.5-9B}} \\
\makecell[l]{\hspace{1.2em}with Original Queries}
& 33.18 & 46.34 & 26.79 & 48.72 & 41.71 & 31.02 & 37.67 & 31.91 & 21.90 & 8.31 & 41.50 & 40.69 & 34.15 \\
\makecell[l]{\hspace{1.2em}with \textbf{QReason}}
& 33.41 & 47.34 & 29.63 & 52.58 & 42.45 & 32.05 & 39.14 & 34.21 & 24.29 & 9.43 & 42.56 & 45.13 & 36.02 ($\uparrow$1.87) \\
\midrule

\multicolumn{14}{l}{\textbf{Qwen2.5-14B-Instruct}} \\
\makecell[l]{\hspace{1.2em}with Original Queries}
& 29.00 & 41.73 & 23.20 & 48.20 & 43.59 & 28.10 & 37.71 & 23.18 & 22.45 & 7.78 & 27.48 & 37.46 & 30.82 \\
\makecell[l]{\hspace{1.2em}with \textbf{QReason}}
& 29.41 & 44.49 & 24.75 & 49.14 & 44.74 & 27.85 & 37.17 & 24.51 & 26.40 & 10.25 & 33.46 & 40.83 & 32.75 ($\uparrow$1.93) \\
\midrule

\multicolumn{14}{l}{\textbf{Qwen3-14B}} \\
\makecell[l]{\hspace{1.2em}with Original Queries}
& 31.97 & 44.15 & 27.21 & 51.79 & 43.25 & 29.19 & 39.07 & 33.07 & 26.17 & 7.93 & 37.69 & 40.03 & 34.29 \\
\makecell[l]{\hspace{1.2em}with \textbf{QReason}}
& 33.24 & 46.36 & 28.06 & 54.34 & 44.30 & 30.77 & 37.75 & 35.93 & 27.07 & 9.67 & 39.37 & 42.21 & 35.76 ($\uparrow$1.47) \\

\bottomrule
\end{tabular}
\label{tab:small_llm_rerankers_bright}
\end{table*}

\begin{table*}[t]
\centering
\footnotesize

\setlength{\tabcolsep}{7pt}
\renewcommand{\arraystretch}{1.2}
\caption{Additional results on FreshStack. The best score is shown in \textbf{bold} and the second best is \underline{underlined}.}
\begin{tabular}{lcccccc}
\toprule
  & LangChain & Yolo & Laravel & Angular & Godot & Avg \\
\midrule
\makecell[l]{Qwen3-Embedding-8B}
& 25.34 & 32.16 & 34.24 & 28.53 & 26.81 & 29.42 \\
\midrule
\makecell[l]{ReasonRank (32B)}
& \underline{31.52} & \textbf{37.11} & 36.73 & 31.78 & \underline{31.24} & \textbf{33.68} \\
\midrule

\multicolumn{7}{l}{\textbf{Qwen2.5-32B-Instruct}} \\
\makecell[l]{\hspace{1.2em}with Original Queries}
& 29.97 & 28.23 & 32.26 & 29.72 & 28.46 & 29.73 \\
\makecell[l]{\hspace{1.2em}with \textbf{QReason}}
& \textbf{31.76} & 30.12 & 33.58 & \underline{32.13} & \textbf{32.22} & 31.96 ($\uparrow$2.23) \\
\midrule

\multicolumn{7}{l}{\textbf{Qwen3-30B-A3B-Instruct-2507}} \\
\makecell[l]{\hspace{1.2em}with Original Queries}
& 25.34 & 32.16 & 34.30 & 28.53 & 26.83 & 29.43 \\
\makecell[l]{\hspace{1.2em}with \textbf{QReason}}
& 27.85 & 34.41 & \underline{37.08} & 29.57 & 30.88 & 31.96 ($\uparrow$2.53) \\
\midrule

\multicolumn{7}{l}{\textbf{Qwen3.5-35B-A3B}} \\
\makecell[l]{\hspace{1.2em}with Original Queries}
& 29.32 & 31.82 & 36.96 & 30.03 & 28.70 & 31.37 \\
\makecell[l]{\hspace{1.2em}with \textbf{QReason}}
& 30.55 & \underline{35.18} & \textbf{39.01} & \textbf{32.99} & 30.41 & \underline{33.63} ($\uparrow$2.26) \\

\bottomrule
\end{tabular}
\label{tab:freshstack_results}
\end{table*}

\begin{table*}[t]
\centering
\footnotesize

\setlength{\tabcolsep}{4.5pt}
\renewcommand{\arraystretch}{1.2}
\caption{Additional results on NeuCLIRBench. The best score is shown in \textbf{bold} and the second best is \underline{underlined}.}
\begin{tabular}{lccccccccc}
\toprule
& \multicolumn{4}{c}{\textbf{Monolingual}}
& \multicolumn{3}{c}{\textbf{Cross-Lingual}}
& \multirow{2}{*}{\textbf{MLIR}}
& \multirow{2}{*}{Avg} \\
\cmidrule(lr){2-5}
\cmidrule(lr){6-8}
& Eng & Fas & Rus & Zho
& Fas & Rus & Zho
& & \\
\midrule

\makecell[l]{BGE-M3 Sparse}
& 31.90 & 36.74 & 28.16 & 27.29
& 4.11 & 5.09 & 8.57
& 5.61 & 18.43 \\
\midrule

\makecell[l]{ReasonRank (32B)}
& 44.53 & 52.67 & 48.32 & 45.19
& 9.11 & 12.14 & 20.05
& 12.82 & 30.60 \\
\midrule

\multicolumn{10}{l}{\textbf{Qwen2.5-32B-Instruct}} \\
\makecell[l]{\hspace{1.2em}with Original Queries}
& 44.08 & 53.02 & 47.39 & 42.13
& 8.28 & 11.10 & 19.01
& 12.79 & 29.73 \\
\makecell[l]{\hspace{1.2em}with \textbf{QReason}}
& 44.19 & 55.21 & \underline{48.90} & \underline{45.84}
& 8.66 & \textbf{12.83} & \textbf{22.15}
& 11.29 & \underline{31.13} ($\uparrow$1.40) \\
\midrule

\multicolumn{10}{l}{\textbf{Qwen3-30B-A3B-Instruct-2507}} \\
\makecell[l]{\hspace{1.2em}with Original Queries}
& \underline{44.83} & 53.90 & 47.57 & 43.78
& 8.36 & 11.39 & 19.46
& 12.60 & 30.24 \\
\makecell[l]{\hspace{1.2em}with \textbf{QReason}}
& \textbf{45.49} & 54.43 & 48.62 & 44.88
& \textbf{9.32} & \underline{12.70} & 20.18
& \underline{13.22} & 31.11 ($\uparrow$0.87) \\
\midrule

\multicolumn{10}{l}{\textbf{Qwen3.5-35B-A3B}} \\
\makecell[l]{\hspace{1.2em}with Original Queries}
& 43.11 & \underline{55.46} & 48.81 & 45.54
& 8.68 & 11.83 & 19.50
& 12.53 & 30.68 \\
\makecell[l]{\hspace{1.2em}with \textbf{QReason}}
& 44.73 & \textbf{56.13} & \textbf{50.28} & \textbf{46.22}
& \underline{9.18} & 12.16 & \underline{20.95}
& \textbf{13.26} & \textbf{31.61} ($\uparrow$0.93) \\

\bottomrule
\end{tabular}
\label{tab:neuclir_results}
\end{table*}

\end{document}